\documentclass[aps,prl,twocolumn,showpacs,a4paper,unsortedaddress,amsmath,amssymb,byrevtex]{revtex4}

\usepackage[latin1]{inputenc}
\usepackage{graphicx}
\usepackage{color}
\usepackage{dcolumn}
\usepackage{bm}

\begin{document}

\title{
Magnetic anisotropies of late transition metal atomic clusters
}

\author{Lucas Fern\'andez-Seivane}
\author{Jaime Ferrer}

\affiliation{Departamento de F\'{\i}sica, Universidad de Oviedo, 33007 Oviedo, Spain}

\date{\today}

\begin{abstract}
We analyze the impact of the magnetic anisotropy on the geometric structure and magnetic
ordering of small atomic clusters of palladium, iridium, platinum and gold, using Density
Functional Theory. Our results highlight the absolute need to include self-consistently
the spin orbit interaction in any simulation of the magnetic properties of small atomic clusters,
and a complete lack of universality in the magnetic anisotropy of small-sized atomic clusters.
\end{abstract}


\maketitle
Nanostructures of all kinds display a wealth of fascinating geometric,
mechanical, electronic, magnetic or optical properties. The exploding field of Nanoscience
pretends to understand, handle and tailor these properties for human benefit.
Atomic clusters and chains, molecular magnets, and a number of organic molecules like
for instance metallocenes indeed show novel magnetic behaviors, that include the
enhancement of magnetic moments due to the reduced coordination and symmetry of the
geometry\cite{Bil94}, and a rich variety of new non-collinear magnetic structures that are absent in bulk
materials\cite{Oda98}. Among all these devices, metallic atomic clusters (MACs)\cite{Bal05} stand out since,
on the one hand they represent the natural bridge between atomic and materials physics
and, in the other, they can be grown, deposited on surfaces or embedded in
diverse matrices, and characterized with relatively well-established techniques. Further,
the magnetic properties of MACs show promise for a wide spectrum of applications, ranging
from medicine to spintronics.

Magnetism in small MACs has been extensively studied both experimentally and theoretically
along the past decade\cite{Hak02,Fer06}.
It is therefore somewhat surprising that, even though spin-orbit effects are also expected to be
enhanced in MACs, there are very few published theoretical papers that include the
spin-orbit interaction (SOI) in their simulations\cite{For99,Bal00,Xia04,Hud06}. Among these,
only a handful have actually looked at the explicit effects of the SOI on the magnetism
of MAC. Pastor and coworkers have studied the magnetic anisotropy (MA)\cite{Pas95}
of clusters of 3d Transition Metal atoms, using a phenomenological tight-binding scheme.
To the best of our knowledge, there are no ab initio studies of the impact of the SOI on the magnetism of
4d and 5d MAC. This includes not only explicit calculations of the magnetic anisotropy energy (MAE),
but also and more
importantly, whether the SOI modifies the ground state and the tower of lowest lying (magnetic) states of
a given cluster.

We present in this article a series of calculations of selected atomic clusters of 4d- and specially
5d-elements, that show that the SOI is a key ingredient in any ab initio simulation of heavy transition
metal MAC, that should not
be overlooked. These include Pd$_n$, Ir$_n$, Pt$_n$ and Au$_n$, with n=2, 3, 4 and 5, and also Pd$_6$
and Au$_6$, Au$_7$. The rationale behind our choice is that 5d MAC are expected to have the largest MAEs;
among all 5d elements, gold is monovalent and should display a simpler behavior; comparing the magnetic
properties of gold MAC with those of its neighboring Pt and Ir elements should shed light on the possible
trends; further, Pd is isoelectronic to Pt and it is important to check whether they display similar magnetic
behavior. We have found a number of interesting results. First, the SOI always alters
slightly the geometry of MAC, therefore generalizing the results obtained for platinum\cite{Hud06} to
other elements. Furthermore, the SOI can lead to qualitative changes in the atomic geometry in specific
cases, as for example Ir$_3$, Ir$_4$ and Au$_7$. Second, the MA generated by the SOI has a strong impact on
the magnetism of the tower of lowest lying states of the clusters. The MA selects orientations
that are typically parallel (PR) or perpendicular (PP) to specific bonds in each cluster. Thus,
MA and tendency to ferromagnetic alignment conspire together for ferromagnetism, or frustrate each
other, depending on the geometrical arrangements of the atoms in each cluster. Hence, the lack of universality
of small MACS is even more evident for their magnetic properties than for their geometry. Finally, atomic
spin and orbital moments align in parallel to each other.

We have used the ab initio Troullier-Martins\cite{Tro91} pseudopotential code SIESTA\cite{SIESTA}.
We have added to this code a fully self-consistent and spin non-collinear implementation of the SOI as
described in detail in Ref. [\onlinecite{Lucas06}]. This implementation has been tested satisfactorily
in several bulk semiconductors and metals\cite{Lucas06}, some molecular magnets\cite{Post06}, a large
number of atomic chains\cite{Lucas07} and a good number of small metallic (in the present work) and
intermetallic clusters\cite{Post06} by now. We have crosschecked our results for
Ir$_2$ and Pt$_{2,3}$ with the plane-wave,
Vanderbilt pseudopotentials code Quantum ESPRESSO\cite{PWSCF}. We have employed both LDA and GGA
approximations \cite{Per81,Per96} for all dimers and trimers to assess the reliability of our results,
except for Ir$_3$.
We have used LDA for larger MACs since LDA provides a slightly more accurate description of bulk Pd, Ir,
Pt and Au, including the magnetism of Pd\cite{Ale06}. We proved recently that the ground state and first
excited isomers of Pd clusters are rather insensitive to the choice of correlation functional and
pseudopotential\cite{Agu06}. We have found here that the MAE further enhances the
consistency of LDA and GGA in Pd, Pt and Ir MACs. In contrast, we confirm that the choice of functional
is important for Au$_n$ clusters\cite{Bas04,Fer06}, even after including the SOI. We have borrowed the
atomic configuration
for the pseudopotentials from Quantum ESPRESSO package's web page\cite{PWSCF}. We have checked that they
produce a correct description for the bulk phases, and have also benchmarked with other choices of matching
radii in the case of some Pd$_n$, Pt$_n$ and Ir$_n$ MACs. We have included non-linear core
corrections for Pd to account for the effect of semicore states in the valence. We have used in all
cases a very complete triple zeta doubly polarized basis set (three pseudo atomic orbital for s and d shells,
and two for the unoccupied p shell), with long radii that provides a very accurate description of the structural
and electronic properties of the corresponding bulk materials. We have explicitly tested and set tolerances in
all parameters (density matrix, electronic temperature, real space grid and forces) to achieve accuracies
in the range 1-3 K. We have calculated the individual atomic spin moments $\vec{S}_i$ using a Mulliken analysis, while
to estimate the atomic orbital moments $\vec{L}_i$ we needed to compute previously the matrix elements of the
orbital angular operator in the chosen pseudo atomic basis. We have taken between two and five different
seeds both for the geometry and for the magnetism of each cluster, making a total of five to
ten different initial configurations for each of the simulated clusters.

We discuss first the highly non-trivial behavior of the magnetic anisotropy of the dimers
Pd$_2$, Ir$_2$ and Pt$_2$ and Au$_2$, that we show in Figs. 1 (a1) through (d1) for LDA. We have found
that GGA agrees with LDA in all the essential details\cite{EPAPS}. Pd$_2$ shows
a weak MAE of about 5 meV that favors an alignment perpendicular to the dimer axis at its equilibrium
distance. This dimer displays a low to high orbital angular momentum transition at longer elongations, which
swaps the MA to PR. In marked contrast to Pd$_2$, the spin moments of the isoelectronic Pt dimer
show PR alignment at its  equilibrium distance. The MAE is estimated as 100 meV or 35 meV, depending on whether
LDA or GGA is used.
The two electrons of a single Pd or Pt atom arrange in a high spin (S=1) configuration, according to Hund's rule.
We find that the impact of the SOI is larger for the Pt$_2$ dimer, where the spin moment is noticeably smaller than
2 $\mu_B$, and different for the two orientations, as shown in Fig. 1 (c2). For Ir$_2$, the SOI also
favors PR alignment, but the predicted MAE at the equilibrium distance of the dimer
is of about 60 meV in the LDA approximation. The spin moment is twice larger than for Pt$_2$ and Pd$_2$
and is also slightly different for the two orientations. We have found that the two electrons in
the gold dimer pair in a singlet state and therefore this clusters does show zero MAE.
Figs. 1 (a3-d3) show the orbital moments of the different dimers.
The atomic orbital moments align parallel to the spin moments always, therefore
spelling out that the $\vec{L}\cdot\vec{S}$ product favors ferromagnetic alignment of these moments.
Notice that the SOI enhances the orbital
magnetism in these dimers, as is reflected by the fact that the orbital moment of Ir$_2$ and Pt$_2$ is similar,
but much larger than that of Pd$_2$.
If we take these results as a rough guide for the anisotropy of bigger clusters, we expect a tendency
to PP alignment for Pd MACs, while Pt and Ir clusters will prefer PR orientations.
Before moving to bigger clusters, we note that
a tendency to PP alignment is easily fulfilled for planar structures, while it leads to frustration
(e.g. non-collinearity) if the geometry is three-dimensional; tendency to PR alignment on the other hand
can only be fulfilled without frustration for linear geometries.

\begin{figure}
\includegraphics[width=0.9\columnwidth,angle=-0]{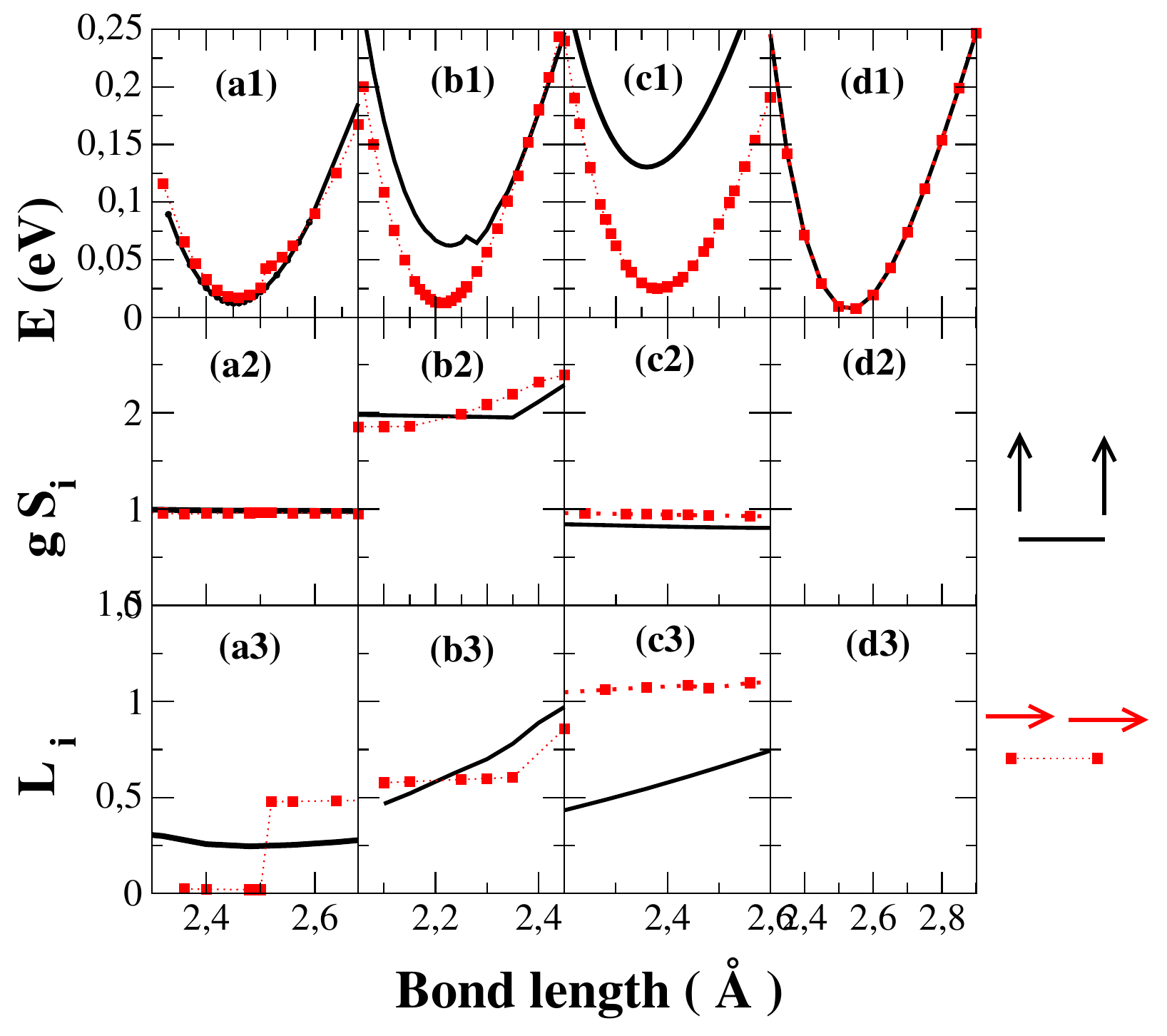}
\caption{(Color online) Calculated ground state energy $E$, spin moment per atom times the gyromagnetic ratio
$g\,S_i$ and orbital moment per atom $L_i$ of (a1-a3) Pd$_2$,
(b1-b3) Ir$_2$, (c1-c3) Pt$_2$ and (d1-d3) Au$_2$,
as a function of inter-atomic distance for spins perpendicular (solid line) or parallel
(dotted line) to the axis of the molecule. A different reference energy has been set for each dimer
to provide for uniformity in the energy values. Figures (d2-d3) demonstrate that the moments of the gold dimer
are exactly zero.}
\end{figure}

We pass now to discuss the geometry and magnetism of the ground state of the trimer MACs, that we show in Fig. 2.
We find that Pd and Pt MACs form a triangle, while Ir arranges in a straight line, and Au$_3$ arranges
as either a
triangle (LDA) or a zigzag (GGA). The tendency to PP alignment of the spin moments in Pd$_2$ leads to
expect out-of-plane alignment for the Pd triangle. But the moments in Pd$_3$ actually align
ferromagnetically in the plane of the triangle, in a fork-like fashion. A closer look to this cluster
reveals that the interatomic distances in the trimer are larger, and actually occur at the crossover
length from PP to PR alignments in the dimer. This leads to a sort of magnetostrictive effect whereby one of
the bonds becomes slightly longer than the other two, hence slightly distorting the
equilateral triangle. The spin moments align perpendicular to that bond. Pt$_3$ is a perfect equilateral
triangle, that does shows no magnetostriction at all. The spin moments in platinum arrange in a fork-like
fashion in the plane of the triangle, as those in Pd$_3$, but now the moments are slightly non-collinear.
Pd$_3$ and Pt$_3$
also share the first excited isomer, which shows out-of-plane anisotropy and ferromagnetic alignment, while
the second one is a vortex for Pd$_3$, and shows out-of-plane antiferromagnetic alignment in the
case of Pt$_3$. A rough estimate of the MAE for these two trimers can be made by taking the energy difference
between the ground state and the first and second excited isomers, since they share the same geometry.
This criterion yields MAEs of 25 and 35 meV for Pd and Pt respectively, if we use the first state, and 60 and
90 meV when we use the second state. Interestingly, these estimates show similar values for the MAE of these two
MACs, despite their different atomic weight.
The PR MA in Ir$_3$ is strong enough to modify its geometry, stabilizing a linear structure instead of the triangle
that is found when the SOI is not included. The first excited state in
Ir$_3$ is a paramagnetic linear chain that has an excitation energy of 7 meV, while the second,
that lies at a much higher energy of about 300 meV is a triangle.
Au$_3$ is weakly magnetic. LDA predicts a triangle with clear PP in-plane anisotropy which leads to a non-collinear
state, while GGA yields a zigzag arrangement, with the spin moments also showing PP in-plane
anisotropy. The first excited state of LDA Au$_3$ is
a triangle that shows out-of-plane PP anisotropy and has a MAE of 8 meV. GGA predicts smaller MAEs for
gold than LDA, of the order of a few Kelvin\cite{EPAPS}.

\begin{figure}
\includegraphics[width=0.9\columnwidth]{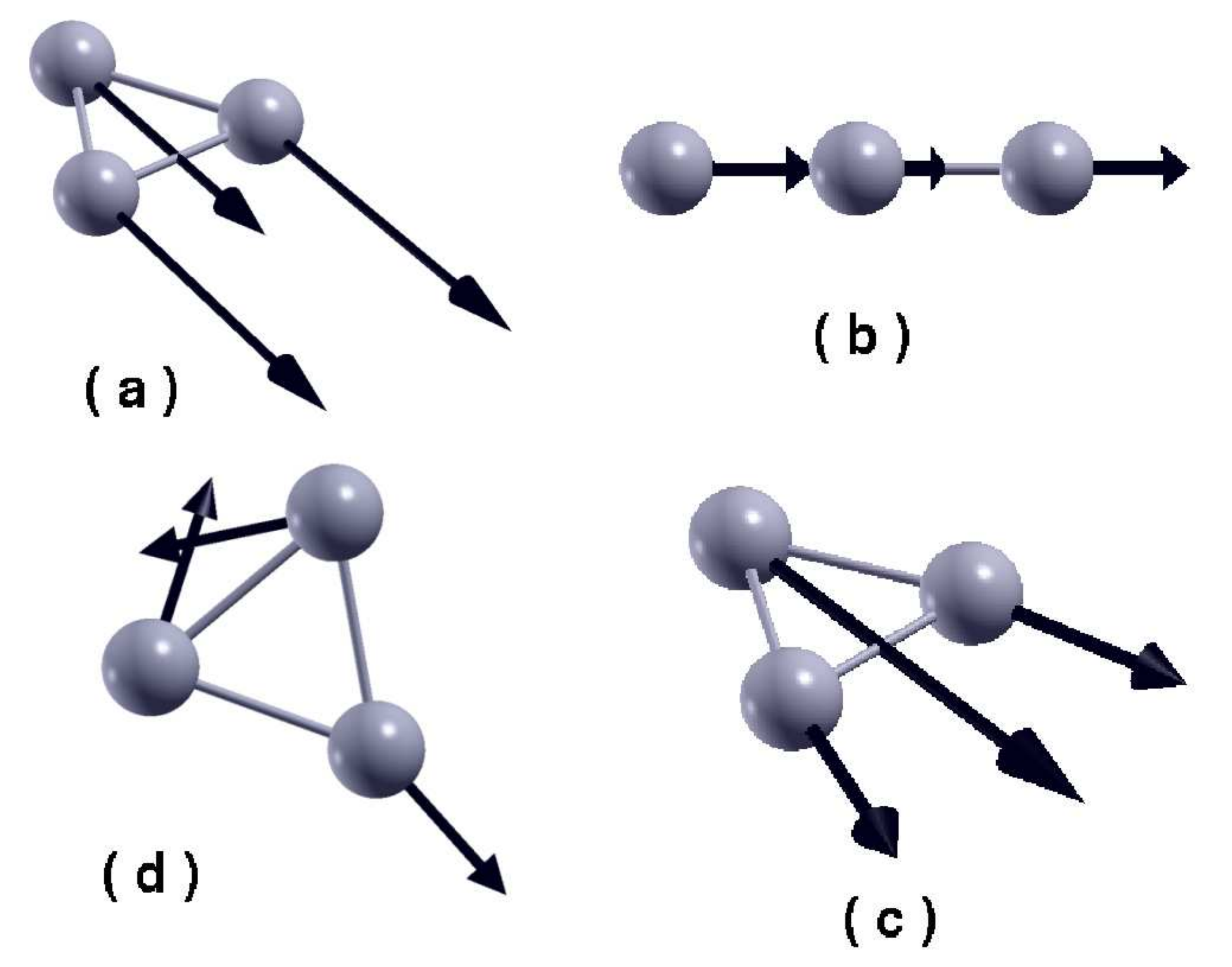}
\vspace{-0.1cm}
\caption{ Equilibrium geometry and spin moments g S$_i$ in the ground state of
the trimers (a) Pd$_3$, (b) Ir$_3$, (c) Pt$_3$ and (d) Au$_3$.
}
\end{figure}

\begin{figure}
\includegraphics[width=0.9\columnwidth]{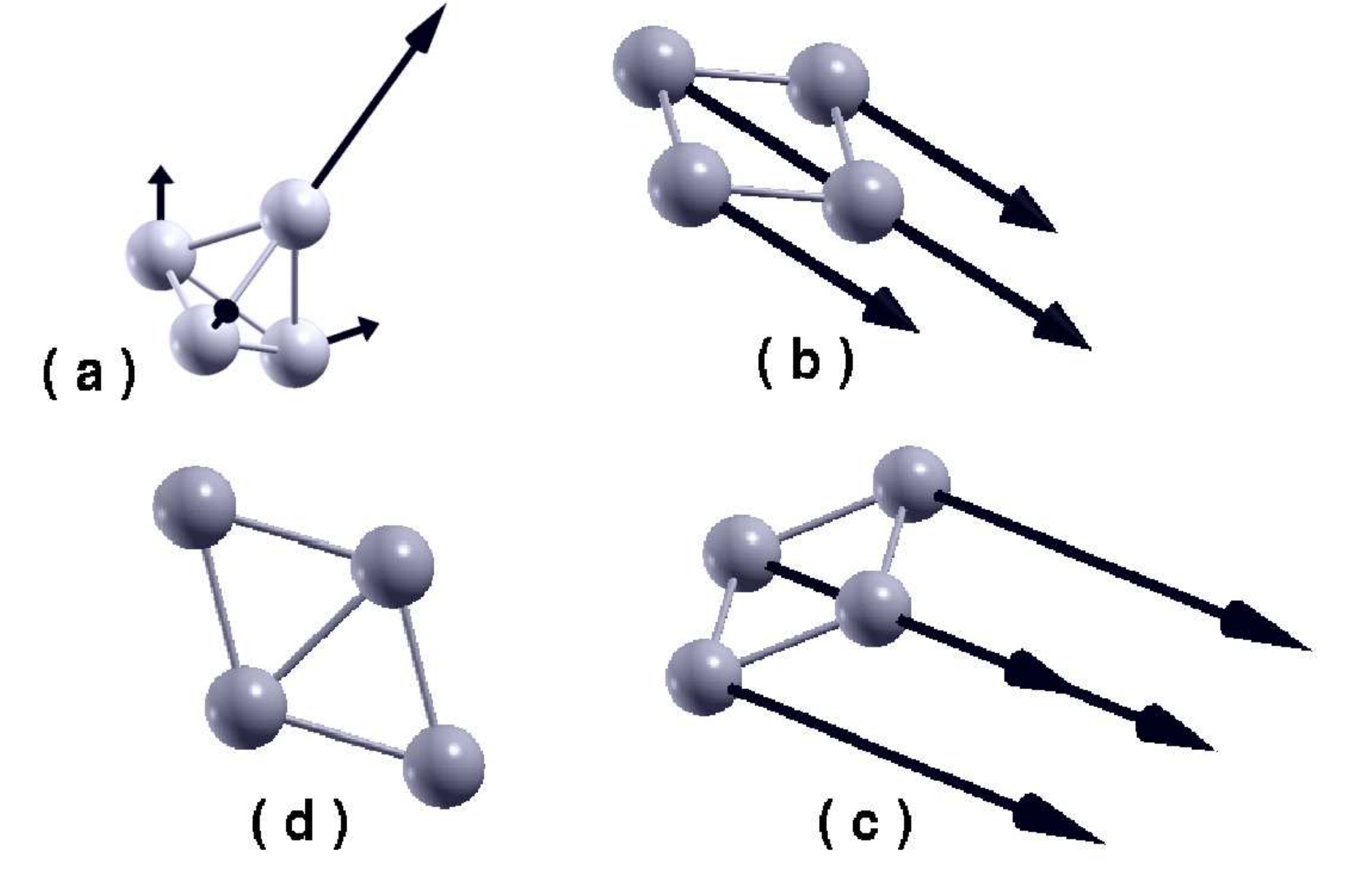}
\vspace{-0.1cm}
\caption{ Equilibrium geometry and spin moments g S$_i$ in the ground state of
the tetramers (a) Pd$_4$, (b) Ir$_4$, (c) Pt$_4$ and (d) Au$_4$.}
\end{figure}

We confirm that the atoms in the ground and first excited states of Pd$_4$ arrange as a tetrahedron\cite{Agu06,EPAPS}. On the contrary, all lowest-lying
states of the fifth-row tetramers show planar geometries, as shown in Fig. 3 for the ground state clusters. Platinum
tetramers form rhombuses, while the lowest lying Ir$_4$ isomers are perfect squares.
An exception is the second excited state of iridium, that is slightly
bent in a sort of saddle. Gold displays a richer variety of lowest lying planar structures.
The tendency to both ferromagnetism and PP MAE in the Pd$_4$ MACs leads to strongly frustrated non-collinear
states, with small energy differences among them. The ground state shows the magnetostrictive effect found in
Pd$_3$, while the magnetic arrangements in the excited states precludes it. The ground state of the iridium and platinum tetramers shows in-plane MAE,
where the atomic spin moments are aligned along one of the diagonals of the square or rhombus.
Their first excited state display out-of-plane MAEs, on the contrary. The excitation energies of these lowest lying states are of about 130 and 50 meV for Ir$_4$ and Pt$_4$, respectively. Ir$_4$ and Pt$_4$ MACs do not
show magnetostrictive effects. The lowest-lying states of gold
clusters are paramagnetic, except for the second one, that shows
a ferromagnetic alignment with out-of-plane MA. The excitation energy of the first gold excited state is of
260 meV\cite{EPAPS}.

The tower of lowest lying states of the palladium, platinum and iridium pentamers show fully 3d arrangements. Palladium
MACs typically form triangular bipyramids. The ground and first excited state of Ir$_5$ are Keops pyramids, while
the second excited state is planar. The ground state of Pt$_5$ is also a Keops pyramid, but the two first excited
states are triangular bipyramids\cite{EPAPS}. Gold pentamers are all planar; the ground state and first
excited isomer displaying an arrangement reminiscent of the Olympics logo. The atomic
spins in all Pd$_5$ states are ferromagnetically aligned and display the same clear tendency to PP MAE as in Pd$_3$ and Pd$_4$. This is to be contrasted to the case of platinum and iridium MACs,
where the tendency to PR MAE can not be satisfied well in these 3d structures. This effect is specially acute for the Keops geometry. The trade-off in this case between spin ferromagnetism
and PR alignments gives rise here to small non-collinearities, as shown in Fig. 4. The spin moments in these two
MACs lie in the basement's plane. They point along its diagonal for Pt$_5$, while for Ir$_5$ the moments are
aligned parallel to the atomic bonds. The ground state of Au$_5$ is magnetic
and displays in-plane MAE. The favored anisotropy is actually PP, which can not be satisfied, giving rise to a
non-collinear arrangement, that we show in Fig. 4. The first excited state of Au$_5$ shows out-of-plane anisotropy,
and has an excitation energy of only 4 Kelvin, revealing again the extreme softness of small gold MACs\cite{EPAPS}.

\begin{figure}
\includegraphics[width=0.9\columnwidth]{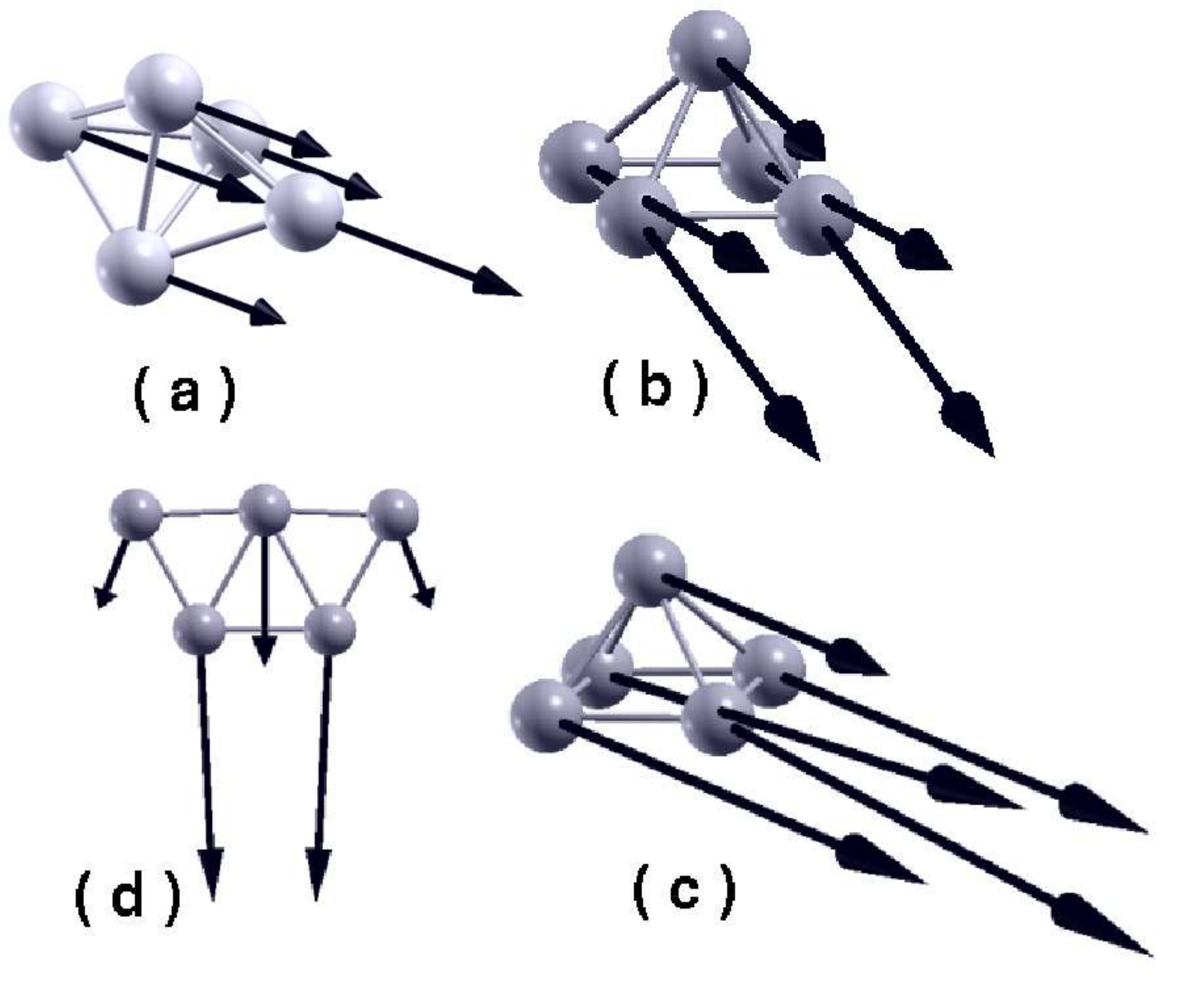}
\vspace{-0.3cm}
\caption{ Equilibrium geometry and spin moments g S$_i$ in the ground state of
the pentamers (a) Pd$_5$, (b) Ir$_5$, (c) Pt$_5$ and (d) Au$_5$.}
\end{figure}

We comment finally on the behavior of a few MACs of larger size. The ground and first excited
states of Pd$_6$ are square bipyramids. The SOI gives rise to small spin and angular moments
in the ground state, that ceases to be paramagnetic. It is the first state, whose excitation energy is of 6 meV, which becomes
paramagnetic. The ground state of Au$_6$ is degenerate. One of the states is a paramagnetic planar triangle
while the other is a square bipyramid, that has small (non-collinear) spin moments, see Figs. 5 (a) and (b). Finally, we
have found that the SOI stabilizes a planar atomic arrangement in Au$_7$. The ground state presents two degenerate
magnetic configurations, with in-plane and out-of-plane MAEs, that we show in Figs. 5 (c) and (d). These again spell out the extreme softness of gold clusters\cite{EPAPS}.

To summarize, we have performed a theoretical analysis of the impact of the spin-orbit interaction in the magnetic
properties of small clusters of Pd, Ir, Pt and Au. We have found that the SOI modifies, slightly always and
qualitatively in a few cases, the geometry of these clusters. More importantly, the spin-orbit interaction gives rise to magnetic anisotropies. These sometimes enhances the tendency to ferromagnetic ordering, while other
compete with it, yielding some non-collinear structures. The MAE also gives rise to an interesting
magnetostrictive effect in Palladium clusters. Our results highlight a lack of universality in the magnetic anisotropy of small-sized atomic clusters.

\begin{figure}
\includegraphics[width=0.9\columnwidth]{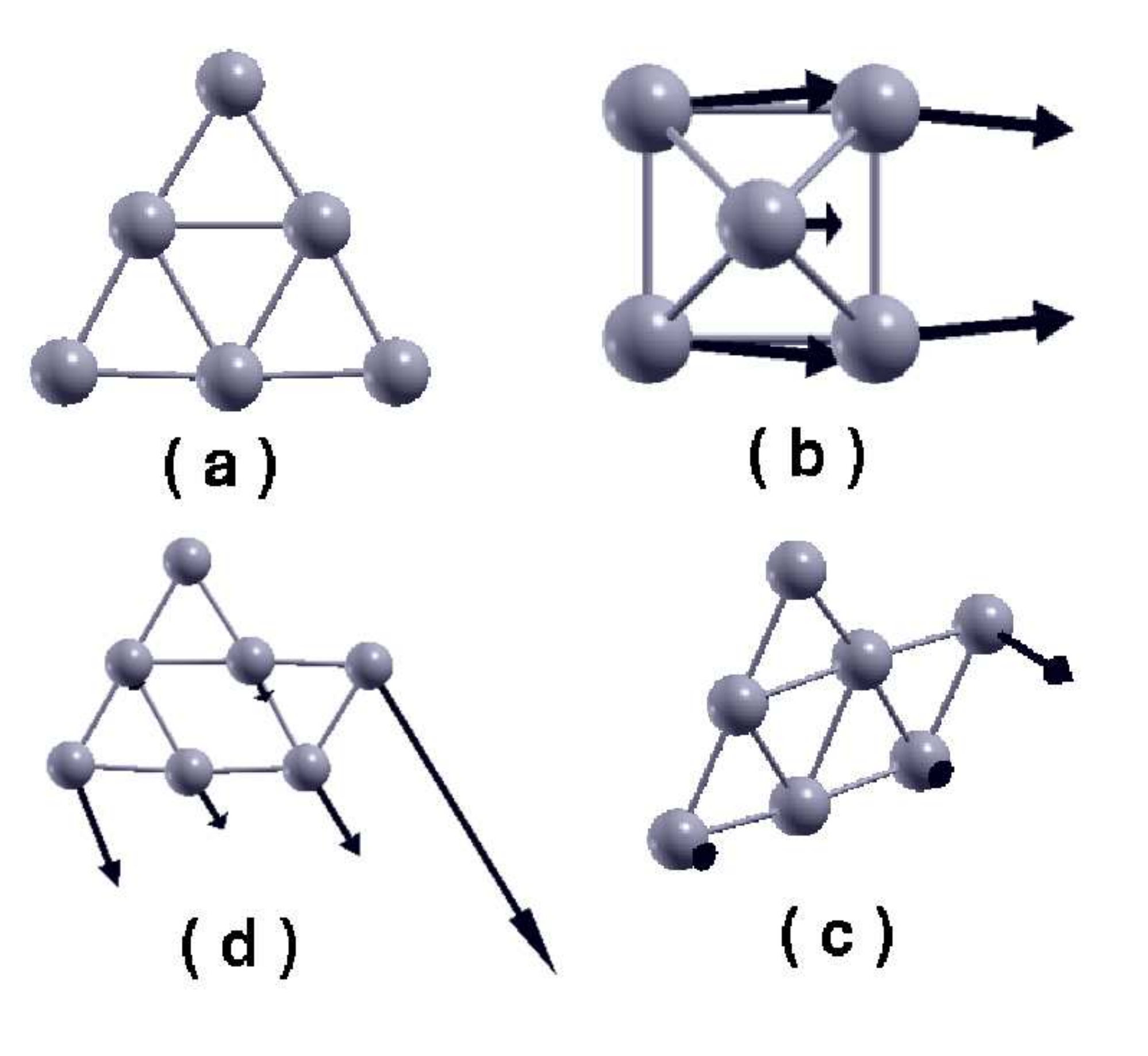}
\vspace{-0.5cm}
\caption{ Equilibrium geometry and atomic spin moments g S$_i$ of the two degenerate ground states of Au$_6$ ((a) and (b))
and Au$_7$ ((c) and (d)).}
\end{figure}

\begin{acknowledgments}
We acknowledge conversations with P. Ordej\'on, J. M. Soler,
D. S\'anchez-Portal, A. Vega, F. Aguilera-Granja,
A. Postnikov, V. M. Garc\'{\i}a-Su\'arez, L. J. Gallego and F. Yndurain,
and financial support from the Spanish government (Project FIS2006-12117-C04-04). LFS is
funded by FICYT (Fellowship BP04-087).
\end{acknowledgments}

\end{document}